\documentstyle[pra,aps]{revtex}
\begin{document}
\draft
\twocolumn[\hsize\textwidth\columnwidth\hsize\csname 
@twocolumnfalse\endcsname
\title{Quantum Distillation Of Position Entanglement With The Polarization Degrees Of Freedom } 
\author{D. P. Caetano and P. H. Souto Ribeiro}
\address{Instituto de F\'{\i}sica, Universidade Federal do Rio de Janeiro, Caixa Postal 68528, Rio de Janeiro, RJ 22945-970, Brazil} 
\date{\today}
\maketitle

\begin{abstract}
Sources of entangled photon pairs using two parametric down-converters
are capable of generating interchangeable entanglement in two different degrees 
of freedom. The connection between these two degrees of freedom allows
the control of the entanglement properties of one, by acting on the other
degree of freedom. We demonstrate experimentally, the quantum distillation
of the position entanglement using polarization analyzers.
\end{abstract}
]
\section{Introduction}
Quantum interference with twin photons from the parametric down-conversion,
has been used in a large number of experiments \cite{1}, concerning the
basic concepts of quantum mechanics and quantum properties of the electromagnetic field. This system has also been shown to be a versatile tool for testing concepts and approaches to be used in quantum information. Among the experiments related to quantum information, we will focus our attention in those dedicated to the production and control of entangled states. Recently, Kwiat and co-workers have introduced a source of polarization  entangled photon pairs, using two nonlinear crystals \cite{2}.
This source consists of two thin crystals close together, so that a very good
spatial mode overlap is obtained between the modes of the two crystals and a
polarization entangled state with high degree of purity and controlled degree
of entanglement is obtained. The same kind of source has been used to demonstrate
the existence of decoherence-free sub-spaces for the polarization, where the
decoherence arose from the coupling with the energy degree of freedom \cite{3}, and reduction of the communication complexity\cite{4}. 

In the quantum information scenario, local operations and classical communication can be used to enhance the quantum correlations of non-maximally entangled states \cite{5}. A scheme to recover the entanglement in a filtering process of the  polarization  degree of freedom was proposed and implemented \cite{6}. This procedure, called {\em quantum distillation}, permits extracting from an ensemble of non-maximally entangled states a sub-ensemble of maximally entangled states with sounding perspectives for quantum information processing. 

In this two crystal source the produced pairs of photons are entangled in other degrees of freedom than polarization. Due to the fact that pairs of photons are generated either in one crystal or in the other they are also entangled in the space of  the transverse components of the momentum. This kind of entanglement is also present for single crystal sources, but when we have two separated  crystals it is possible to describe the state of the field as a superposition of discrete states labeled by the photon pair's birth position, which is the position of each crystal\cite{7,8}. We call it {\em position} entanglement. 
In order to be able to detect the quantum interference arising from the entanglement in this degree of freedom, it is necessary to arrange the crystals in a suitable configuration, so that the phase difference between the states of the superposition can be varied\cite{7,8}.
In the experiments of Refs.\cite{2,3,4} the pairs of photons are distinguishable by
their polarization, but even if they had the same polarization, the observation of the interference due to the position entanglement would not be possible. There, the crystals were so close and so thin that the phase difference between the states of the superposition varies too much slowly with the control parameters.
In this work, we propose and implement a scheme for performing the quantum distillation based on the idea that actions in one degree of freedom affects the entanglement
properties in the other. Using a suitable two crystals configuration to produce photon pairs with interchangeable entanglement in polarization and position, we will demonstrate the distillation of 
position entangled states acting in the polarization  degree of freedom.

\section{Distillation Scheme}
\begin{figure}[h]
\vspace*{7.5cm}
\special{eps: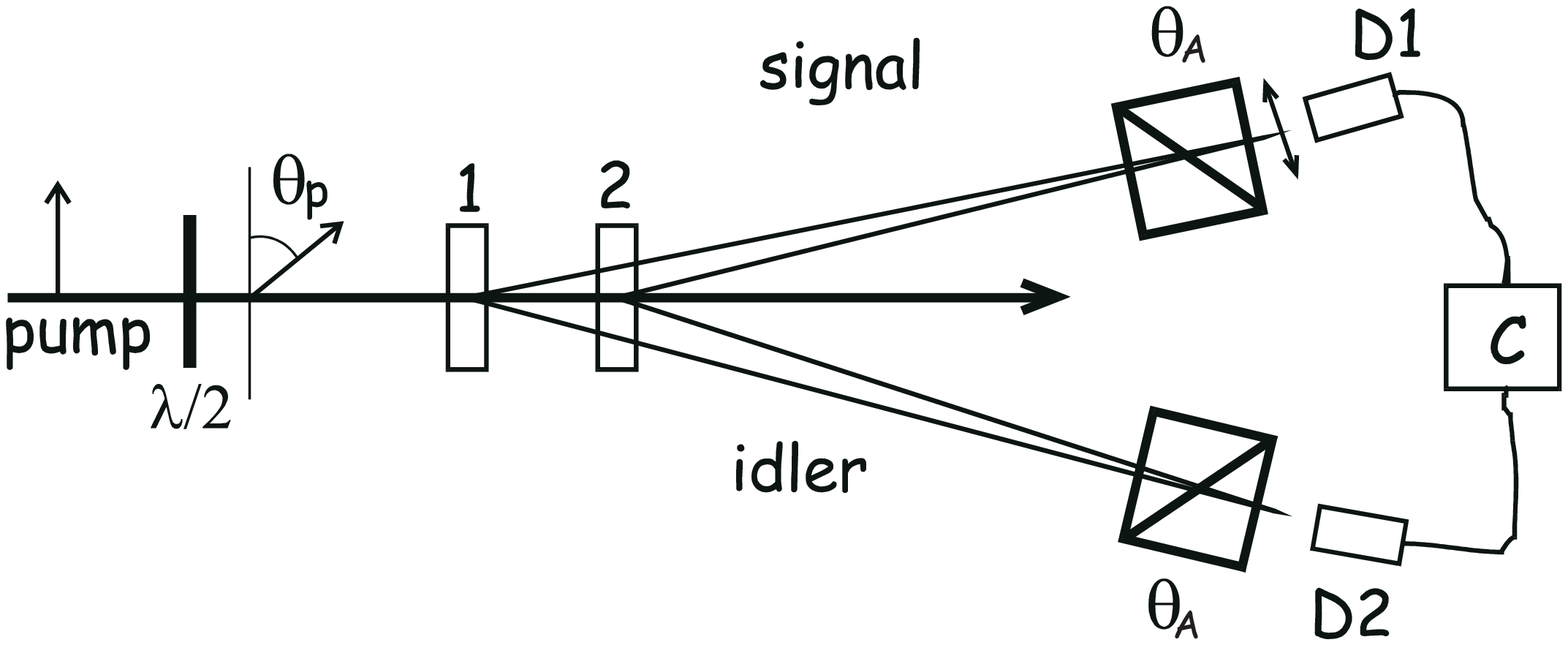 x=8.5cm y=6cm}
\caption{Experimental setup.}
\label{fig1}
\end{figure}

Let us consider the experimental set-up sketched in the Fig. \ref{fig1}. 
A $45^o$ linearly polarized laser beam  pumps two type I nonlinear crystals with  
optical axes perpendicularly oriented, one in the vertical direction(V) and the
other horizontal(H). By parametric down-conversion, two pairs of photons can be generated in two light cones, in each one of the crystals. The first crystal is pumped by the vertical component of the laser, emitting horizontally polarized pairs of photons, while the second one is pumped by the horizontal component producing vertically polarized pairs of photons.
By superposing signal and idler beams coming from each crystal, so that 
spatial mode matching is obtained, we get the position/polarization entangled 
state: 
\begin{equation}
\label{eq1}
|\psi \rangle =\left[ |2_H, \text{0}_V\rangle +e^{i\phi }|0_H,2_V\rangle \right] /\sqrt{2}, 
\end{equation}
where the phase $\phi$ depends on the pump phase at each crystal and on the optical path from crystals to detectors\cite{7,8}. 
This is a maximally entangled state, as both crystals are equally pumped and have
equal probabilities to produce a pair of photons.

Our measurements are focusing
on the degree of entanglement, which is the parameter to be distillated. Note that
the degree of entanglement is the same for both degrees of freedom, however
projections on one sub-space may  change the degree of entanglement differently
for the different degrees of freedom. 
If the two crystals were producing beams with the same polarization, 
the position entanglement could be observed by detecting signal and
idler photons in coincidence. The displacement of the detectors would
vary the optical path differences between the two possible realizations
of the two photon field, giving rise to oscillations in the coincidence counting
rate\cite{7,8}. As the crystals are emitting pairs of photons with different
polarizations, they become distinguishable and the position interference
can only be observed, if polarization analyzers are placed before detectors.
Setting the polarization analyzers to $45^o$, the field after them are not
entangled in polarization anymore, but the position interference can be
observed. 
By changing the pump laser polarization we can control the
degree of entanglement  in such a way that the state of field can be
described by: 
\begin{equation}
\label{eq2}
|\psi \rangle =\cos \theta _P|2_H,0_V\rangle +\sin \theta _Pe^{i\phi }
|0_H,2_V\rangle ,
\end{equation}
where $\theta _P$ gives the pump laser polarization angle.
After crossing the polarization analyzers, with axis oriented both in $\theta_A$
direction, the state the state of the down-converted field is reduced to:

\begin{equation}
\label{eq3}
|\psi \rangle =\cos \theta_P \cos^2\theta_A|2,0\rangle +\sin \theta _P\sin^2 \theta_A 
e^{i\phi }|0,2\rangle .
\end{equation}

>From Eq.\ref{eq3} above, it is seen that by properly orienting the polarization
analyzers, it is possible to control the coefficients of the superposition
and therefore to control the degree of entanglement of the outgoing state.
In order to demonstrate the operation of the scheme, we will change the
pump beam polarization, preparing position entangled states with
different degrees of entanglement. For each input degree of entanglement,
the analyzers will be properly set for recovering the maximal degree of entanglement.

\section{Experiment and Results}
\begin{figure}[h]
\vspace*{7.5cm}
\special{eps: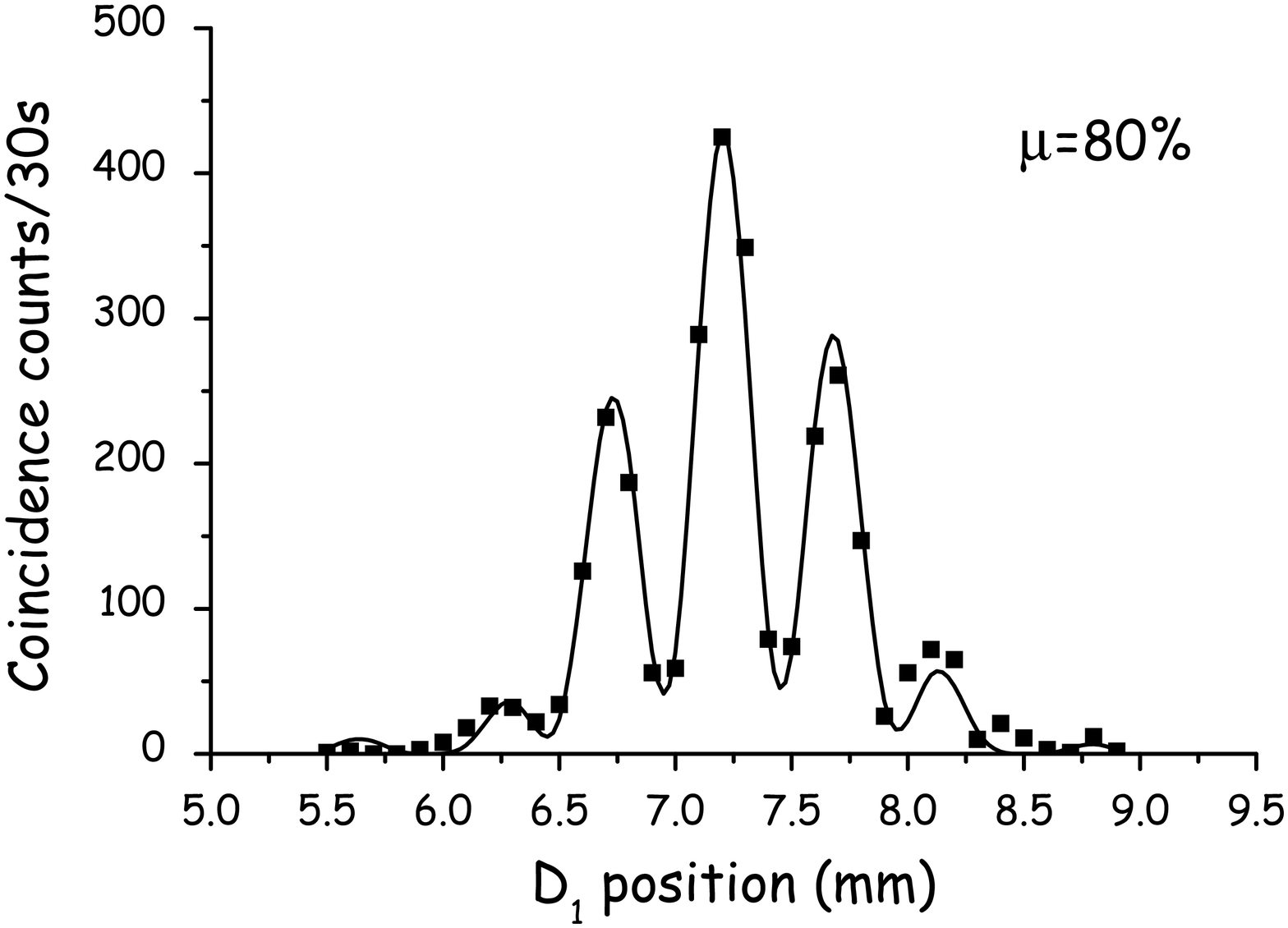 x=8.5cm y=6cm}
\caption{Fourth-order interference fringes measured by transverse scan of the D1 detector keeping D2 fixed. Maximally entangled state. $\mu$ is the visibility of the fringes.}
\label{fig2}
\end{figure}

\begin{figure}[h]
\vspace*{7.5cm}
\special{eps: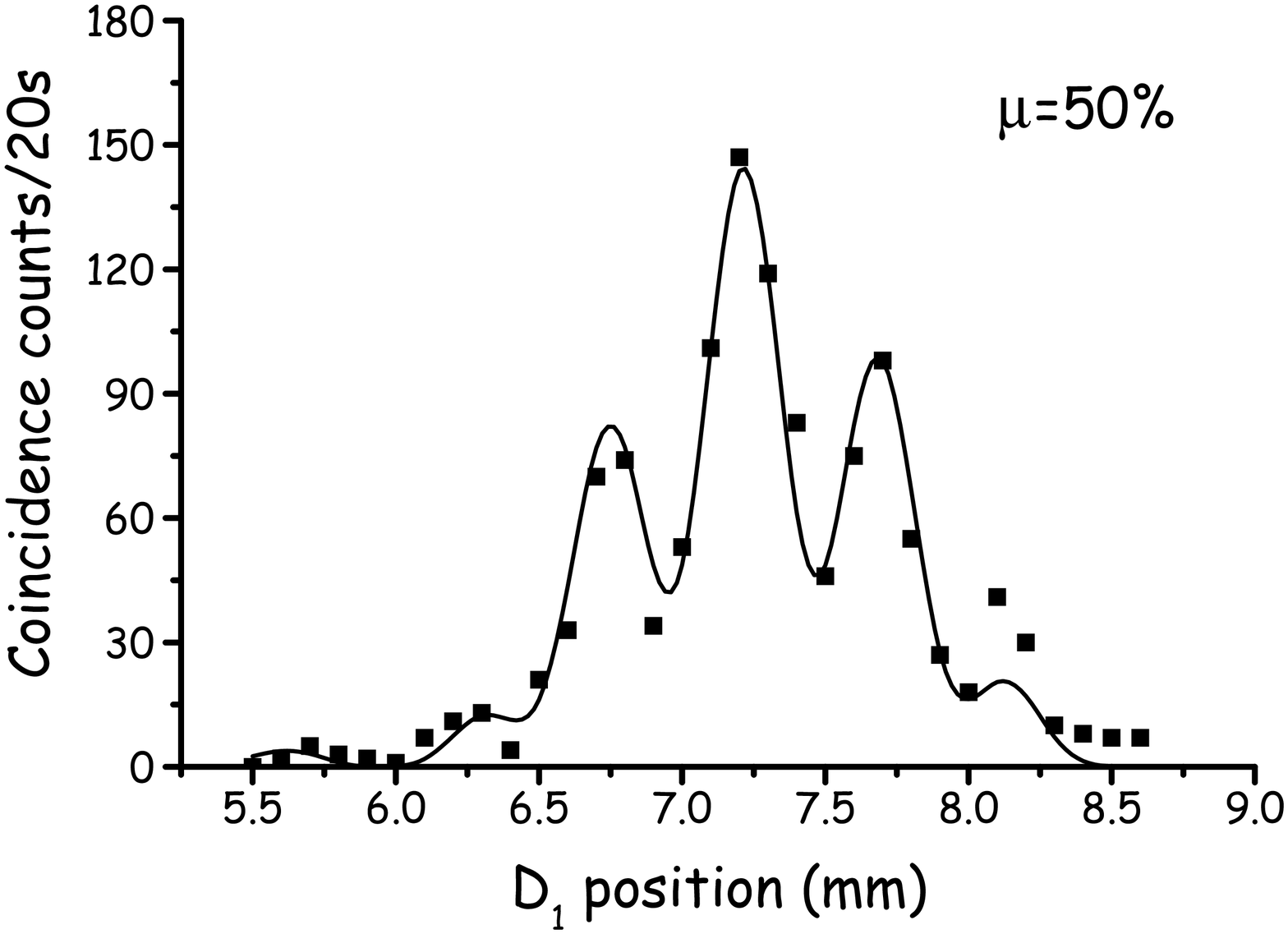 x=8.5cm y=6cm}
\caption{Fourth-order interference fringes measured by transverse scan of the D1 detector keeping D2 fixed. Non maximally entangled state. $\mu$ is the visibility of the fringes.}
\label{fig3}
\end{figure}

\begin{figure}[h]
\vspace*{7.5cm}
\special{eps: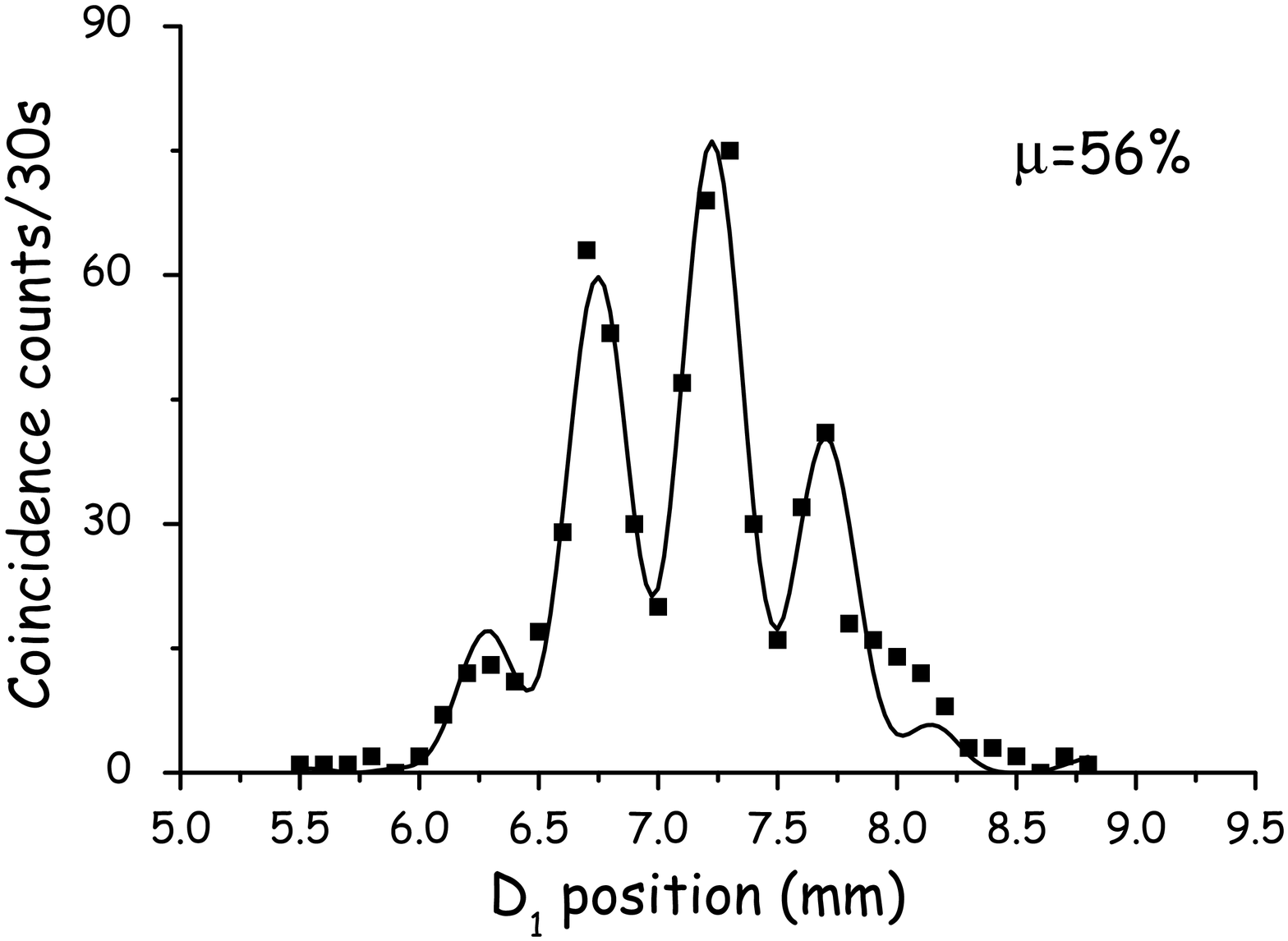 x=8.5cm y=6cm}
\caption{Fourth-order interference fringes measured by transverse scan of the D1 detector keeping D2 fixed. Enhancement of the degree of entanglement by setting the angle of the analyzers to $22.5^o$. $\mu$ is the visibility of the fringes.}
\label{fig4}
\end{figure}

\begin{figure}[h]
\vspace*{7.5cm}
\special{eps: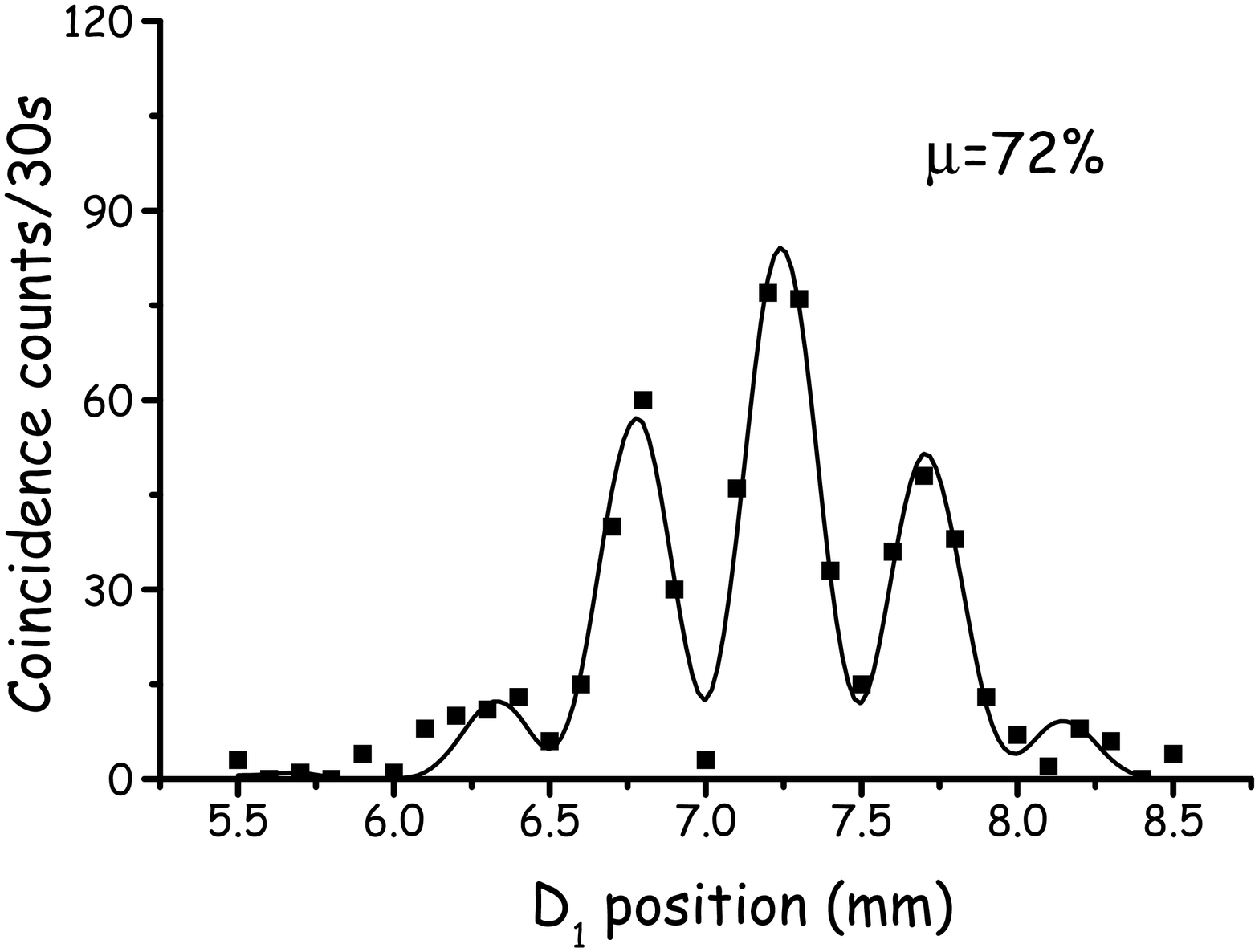 x=8.5cm y=6cm}
\caption{Fourth-order interference fringes measured by transverse scan of the D1 detector keeping D2 fixed. Enhancement of the degree of entanglement by setting the angle of the analyzers to $55^o$. $\mu$ is the visibility of the fringes.}
\label{fig5}
\end{figure}

\begin{figure}[h]
\vspace*{7.5cm}
\special{eps: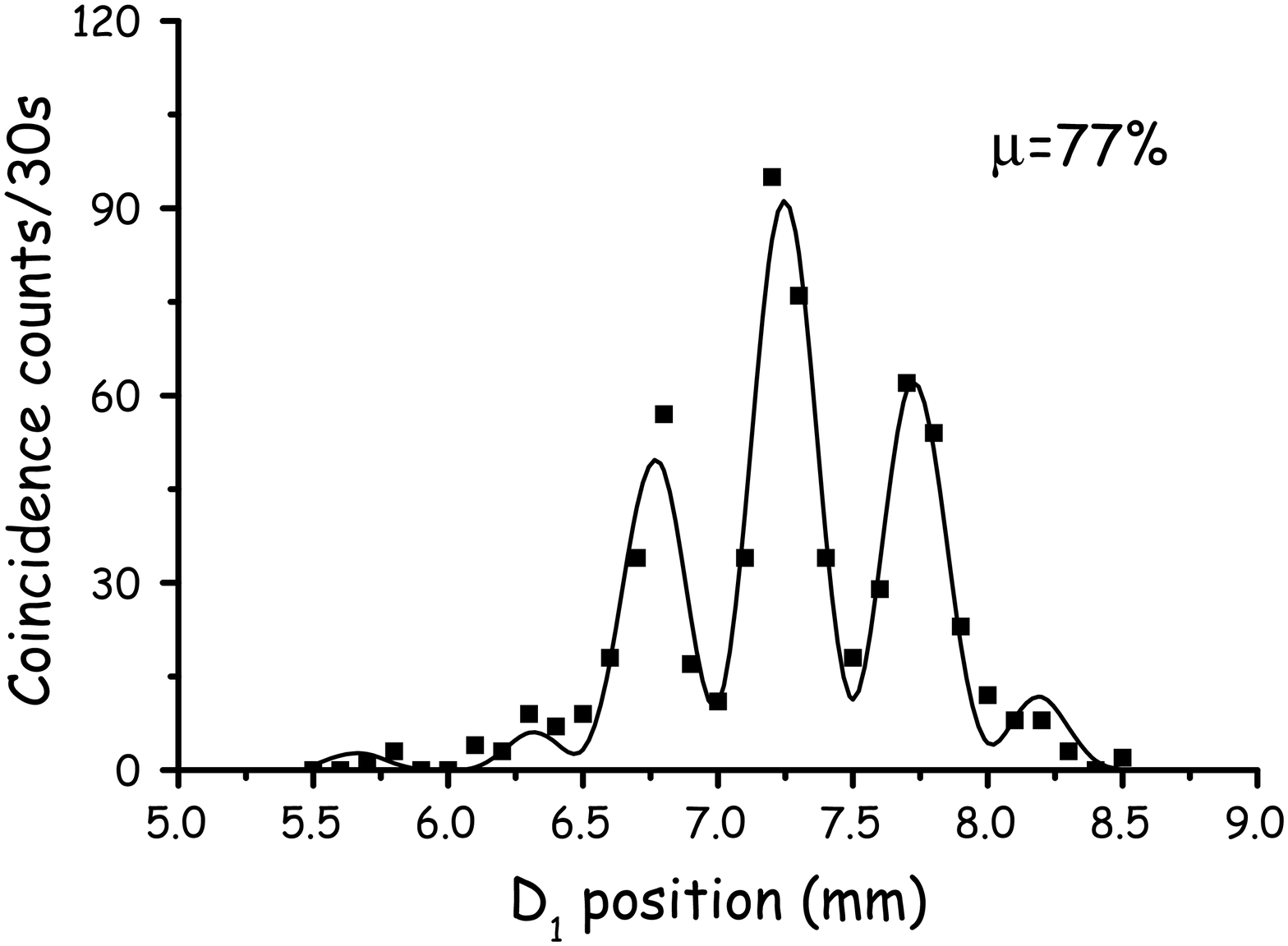 x=8.5cm y=6cm}
\caption{Fourth-order interference fringes measured by transverse scan of the D1 detector keeping D2 fixed. Enhancement of the degree of entanglement by setting the angle of the analyzers to $57^o$. $\mu$ is the visibility of the fringes.}
\label{fig6}
\end{figure}

\begin{figure}[h]
\vspace*{7.5cm}
\special{eps: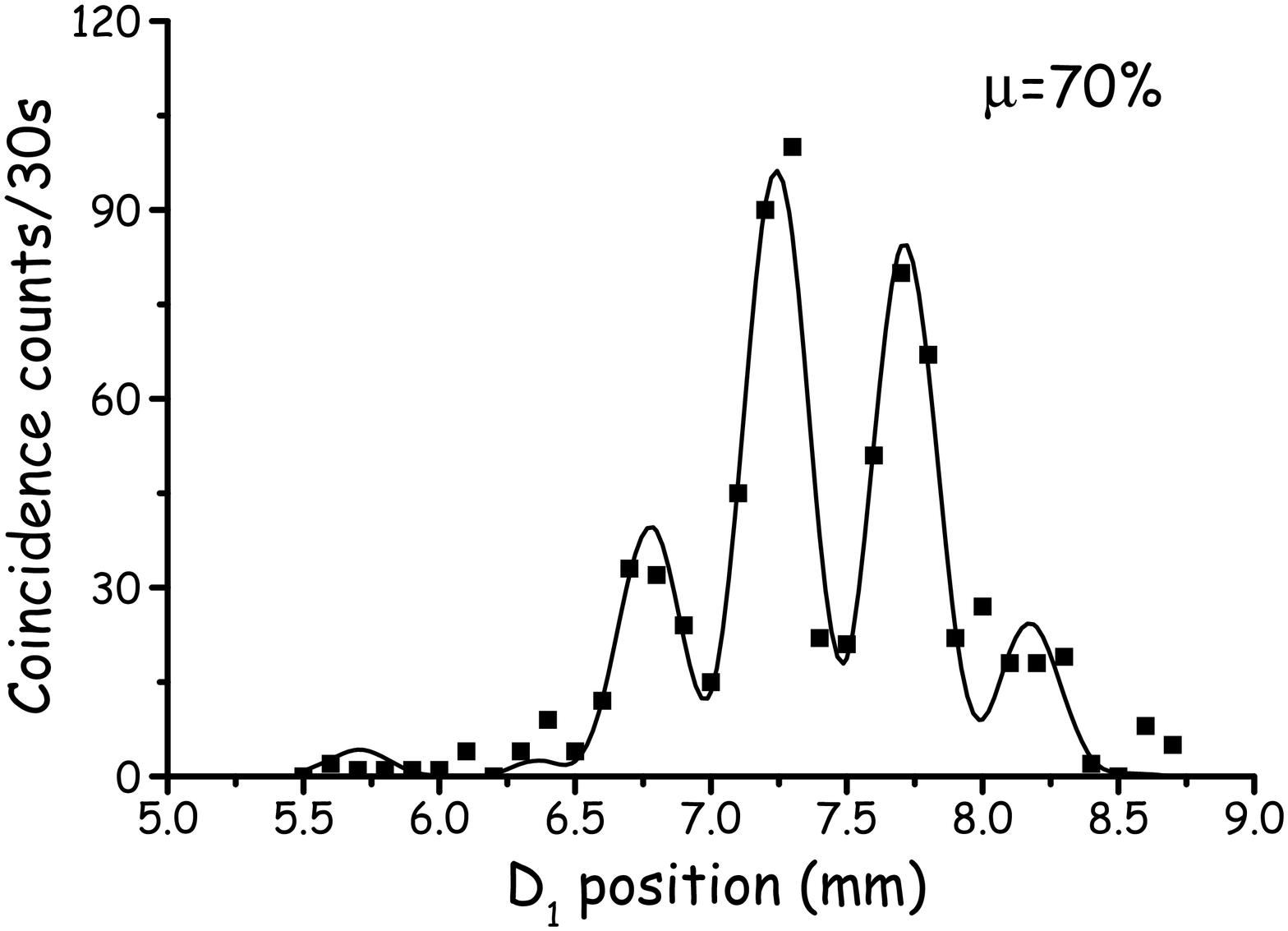 x=8.5cm y=6cm}
\caption{Fourth-order interference fringes measured by transverse scan of the D1 detector keeping D2 fixed. Enhancement of the degree of entanglement by setting the angle of the analyzers to $59^o$. $\mu$ is the visibility of the fringes.}
\label{fig7}
\end{figure}

The experimental setup is shown in Fig. \ref{fig1}. A cw He-Cd laser operating
at $442nm$ pumps two $1cm$ long $LiIO_3$ crystals cut for type I phase
matching. Two signal and idler beams at the degenerate wavelength $884nm$
emerge from the crystals at about $3^o$ with respect to the pump beam direction of
propagation. Crystal 2 is placed about $5mm$ from crystal 1. After crossing the analyzers
the photon pair is
detected with avalanche photodiode counting modules, placed about $80cm$
from crystal 2. The detected light is collected by a thin slit of about 
$0.5mm$, a $10nm$ bandwidth IF(interference filter) centered at $884nm$ 
and a $50mm$ focal length lens at the detector entrance.
The transverse momentum correlations are measured displacing $D_1$ along
the horizontal direction and fixing the analyzers at $45^o$. As a result
fourth-order interference fringes are obtained as displayed in Fig. \ref{fig2},
presenting fringes visibility of about 80\%. This visibility is the signature of the maximally entangled state in our experimental configuration.
The non maximally entangled state is produced by rotating the pump HWP so that  $\theta_P = 22.5^o$. The measurement of the 
fourth-order interference is repeated.
Fig. \ref{fig3} shows the interference fringes with visibility of about 50\%, and 
therefore a lower degree of entanglement. Now we
follow to distill the position entangled state. First we
rotate both analyzers by the same angle as the pump HWP. The interference
pattern visibility is about 56\% as we can see in Fig. \ref{fig4}. In order to find the angles
for the analyzers that maximize the distillation, we have rotated both
analyzers and searched for the maximum visibility. This have been done for the
angles $\theta_A = 55^o$, $\theta_A = 57^o$ and $\theta_A = 59^o$. The results are
displayed in Figs. \ref{fig5}, \ref{fig6}, and \ref{fig7}, with visibilities of about 72\%,
77\% and 70\% respectively. The higher visibility was found for $\theta_A = 57^o$
which corresponds to the analyzers' angle that balance the coefficients of the state predicted by Eq.\ref{eq3}:

\begin{equation}
\label{eq4}
cos\theta_P\cos^2\theta_A$ = $\sin\theta_P\sin^2\theta_A.
\end{equation}
This relation express how the analyzers have to be oriented to distill a non maximally entangled state.

\begin{figure}[h]
\vspace*{14.5cm}
\special{eps: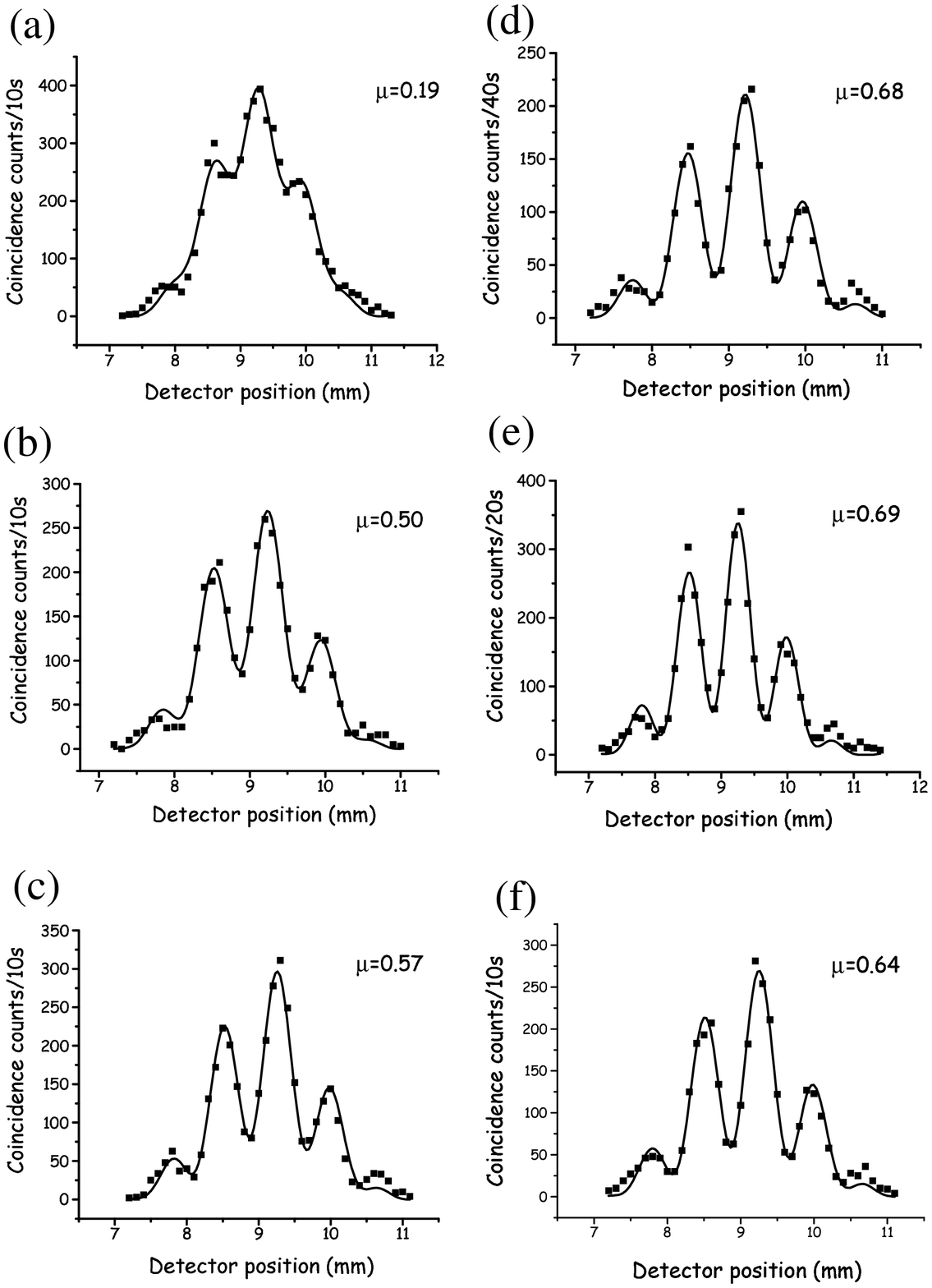 x=8.5cm y=12cm}
\caption{Left side: non maximally entangled states produced by setting the angle of the pump HWP to (a) $\theta _P=10^o$, (b) $\theta _P=20^o$ and (c) $\theta _P=30^o$; right side: enhancement of the degree of entanglement by setting the angle of the analyzers to (d) $\theta _{A}=67^o$, (e) $\theta _{A}=59^o$ and (f) $\theta _{A}=53^o$. $\mu$ is the visilibity of the fringes.}
\label{fig8}
\end{figure}

In order to verify our distillation scheme we prepared entangled states with different degrees of entanglement. Fig. \ref{fig8} shows the application of Eq.\ref{eq4} for several pump laser polarization angles and analyzers properly oriented. The results show the interference patterns for the non maximally entangled states with both analyzers at $45^o$ (left side) and with proper analyzers' angles (right side). In all cases, the degree of entanglement was increased in agreement with Eq. \ref{eq4}. In this set of data, the visibility of the maximally entangled state was about 66\%.

\section{Discussion}

The experimental results presented above are a simple demonstration of how
one can manipulate the entanglement properties of a photon pair, with the use
of interchangeable entanglement in two different degrees of freedom. In our case, the position entanglement can be manipulated acting in the polarization degree of freedom in order to control the degree of entanglement. 

The visibilities of the interference patterns for the maximally entangled states are not 100\%. This is due to technical imperfections. Detection through a very small aperture, would project
both modes (coming from crystals 1 and 2) onto the same transverse mode. 
In the experiment, however, detection
through finite apertures implies in the increase of the degree of  distinguishability of the modes, and
a lower visibility. The simple theory we have used is enough to explain the main features
of the distillation process. However a multimode theory is necessary to take into account
the effects that reduce the maximal visibility.

Envisaging applications in quantum communications, this scheme could be used
for recovering maximal entanglement when a transmitted signal containing information
in the quantum superposition state, is degraded by propagation. In our demonstration,
the degree of entanglement was previously known, as it was prepared by setting the pump HWP
orientation. However, it works for an unknown input degree of entanglement. The receiver
should properly tune the angles of the analyzers for maximizing the fringes visibility.
This action would be effective for restoring the maximal degree of entanglement in the
case where the propagation disturbances change in a time scale much smaller than those
of the communication protocol. The reliability of the position entanglement for 
quantum communication still needs to be demonstrated. We believe that this degree
of freedom may be quite useful, for example, if the entanglement properties are
preserved after propagation through a multimode optical fiber.

Another interesting perspective is to add a third entangled degree of freedom, for
example the entanglement in the orbital angular momentum. This third degree of
freedom could open new possibilities for monitoring and controlling the entanglement
properties of other degrees of freedom.

\section{Conclusion}

We propose and demonstrate experimentally a scheme for performing quantum
distillation of the position entanglement by manipulating the polarization degree of
freedom. 

\section{ACKNOWLEDGMENT}
The authors acknowledge financial support from the Brazillian 
agencies CNPq, CAPES, PRONEX, FAPERJ and FUJB.

\end{document}